# Magnetic and dielectric behavior of the spin-chain compound Er$_2$BaNiO$_5$ well below its Néel temperature

Tathamay Basu[1], Kiran Singh[1,#], N. Mohapatra[2], and E.V. Sampathkumaran[1]

[1]Tata Institute of Fundamental Research, Homi Bhabha Road, Colaba, Mumbai-400005, India

[2]School of Basic Sciences, Indian Institute of Technology Bhubaneshwar, Bhubaneshwar-751013, India

## Abstract

We have recently reported that the Haldane spin-chain system, Er$_2$BaNiO$_5$, undergoing antiferromagnetic order below ($T_N$= ) 32 K, is characterized by the onset of ferroelectricity near 60K due to magnetoelectric coupling induced by short-range magnetic-order within spin-chains. We have carried out additional magnetic and dielectric studies to understand the properties well below $T_N$. We emphasize here on the following: (i) A strong frequency dependent behaviors of ac magnetic susceptibility and complex dielectric properties have been observed at much lower temperatures (< 8 K), that is, *'reentrant multiglass-like'* phenomenon, naturally suggesting the existence of an additional transition well below $T_N$.  ii) *'Magnetoelectric phase coexistence'* is observed at very low temperature (e.g., $T$ =2K), where the high-field magnetoelectric phase is partially arrested on returning to zero magnetic field after a cycling through metamagnetic transition.

[#]Current address: UGC-DAE Consortium for Scientific Research, University Campus, Khandwa Road, Indore - 452001, India

## INTRODUCTION

The Haldane [1] spin-chain systems of the type, $R_2BaNiO_5$ (R= Rare-earths), have been paid a lot of attention for the past few decades from magnetism angle. Recently, some members of this family (R= Gd, Dy, Ho and Er) of oxides attracted further interest from the angle of magnetoelectric (ME) / magnetodielectric (MDE) coupling phenomenon [2-6]. A systematic study on the Er member [5] suggested that a displacive-type mechanism involving magnetic ions is responsible for the onset ferroelectricity well above its Néel temperature ($T_N$= 32 K) and that short-range magnetic correlations are interestingly sufficient enough to induce the same. In the case of Dy member, it was noted that, in addition to the multiferroic behavior setting in at ($T_N$=) 58 K, there is an additional transition at temperatures ($T$) well below $T_N$, as though there is a 'reentrant phenomenon' into a complex multiglass-like state below 10 K. Such a phenomenon is not so commonly known particularly among undoped stoichiometric compounds in the field of multiferroics, barring a few examples (see, Ref. 5, 7 and articles cited therein), though the concept of multi-glass was originally discussed in some doped materials [8]. It is therefore of interest to study low temperature properties of other members of this family well below their respective $T_N$. With this primary motivation, in this article, we report the results of our investigation on the Er member well below its $T_N$. The results reveal that there are glassy anomalies both in magnetization and dielectric properties far below $T_N$, characterizing that this compound also belongs to a new family of 'reentrant' multiferroic oxides. Additionally, there are features in the isothermal data attributable to 'magnetoelectric phase-co-existence' phenomenon.

## EXPERIMENTAL DETAILS

Polycrystalline form of the compound, $Er_2BaNiO_5$, employed in the present investigation is the same as that in Ref. 5. The *ac* susceptibility ($\chi$) measurements with various frequencies ($\nu$) with an ac field of 1 Oe were performed as a function of $T$ with the help of a commercial superconducting quantum interference device (Quantum Design (QD), USA) magnetometer. In addition, *dc* magnetization ($M$) data were obtained with the help of a commercial (QD) vibrating sample magnetometer. The complex dielectric permittivity measurements with different $\nu$ were performed as described in Ref. 5. All the $T$-dependent studies were carried out in the presence of external *dc* magnetic fields ($H$) as well. Unless otherwise stated, all the measurements were performed during warming after cooling the specimen in the absence of an external magnetic-field from the paramagnetic state to the desired temperature. Further experimental details can be obtained from Ref. 5.

## RESULTS AND DISCUSSIONS
**Magnetic and dielectric behavior and multi-glass-like behavior:**

Though $T$-dependent dc $\chi$ behavior is available in the past literature [9], for the sake of characterization, we report our results obtained in a field of 100 Oe for zero-field-cooled (ZFC) and field-cooled (FC) conditions in figure 1. The Curie-Weiss temperature and the effective moment obtained from the linear region in the paramagnetic state are about -11 K (with the negative sign consistent with antiferromagnetic interaction) and 14$\mu_B$ per formula unit respectively. We derive that the moment on Ni is ~ 3.5 $\mu_B$/Ni considering the theoretical moment for Er (~ 9.58 $\mu_B$/$Er^{3+}$). The feature at $T_N$ is so feeble that this can be observed in the derivative curve only (see inset (a) in figure 1). There is a well-defined broad peak at a lower temperature range, that is, around 16 K. A careful inspection of the data in the mainframe of figure 1 reveals that there is an upturn in $\chi(T)$ with decreasing temperature near 4 K with a tendency for the bifurcation of zero-field-cooled (ZFC) and field-cooled (FC) curves. At the same temperature, heat-capacity ($C$) data exhibits a shoulder [5]. Similar behavior was observed in the data measured in 5 kOe (not shown here). This observation indicates that there is another anomaly around this temperature. At this moment, it is worth recalling the evolution process of spontaneous magnetic moments from Er and Ni below $T_N$, as revealed by neutron diffraction results [10, 11]. Though spontaneous moment appears for both Er and Ni at the onset of magnetic ordering, the $T$-dependencies are markedly different; that is, the increase of the Er magnetic moment with decreasing temperature was reported to be rather large for Er, whereas the one on Ni tends to increase only marginally below 30 K. A point to be inferred from the previous neutron diffraction results is that there is no additional well-defined magnetic transition at temperatures well below 32 K; however, for the magnetic moment on Er, there is a change of curvature over a wide temperature range around 15- 20 K, followed by a tendency to saturate below 5 K. Clearly, therefore, the anomalies observed in $\chi(T)$ (as well as in heat-capacity [5]) around 16 K should be associated with these broad thermal variations of the magnetic moment on Er, and not to any magnetic transition. It is argued that the 16K-anomalies arise from the splitting of the Kramer's doublet of trivalent Er by the exchange interaction, followed by depopulation with decreasing temperature [12]. However, the feature below 4 K could arise from another subtle magnetic phase transition, and there is also an evidence from Young's modulus reported by Chepurko et al [13].

In order to get a better insight into the nature of magnetic state below 4 K, we have also measured isothermal remnant magnetization ($M_{IRM}$) behavior at various temperatures. For this purpose, the specimen was zero-field-cooled from 300 K to the desired *T*, then a *dc* field of 5 kOe was applied and kept for 10 min; subsequently, the $M_{IRM}$ was measured as a function of time (*t*) in zero-field. We could see a decay of $M_{IRM}$ with time at 2 K (see inset *b* in figure 1) as expected for spin-glass freezing [14]. The curve was fitted to a stretched exponential form ($[M_{IRM}(t)/M_{IRM}(0)] = a + b \exp[(t/\tau)^{0.5}]$, where a and b here are coefficients, τ is the relaxation time). Interestingly, no such magnetic relaxation behavior was observed at T=5 K and above. This result suggests that there is a difference in the magnetic states below and above ~ 5K.

We have measured *ac* χ response for various values of ν in the magnetically ordered state (see figure 2) and the features in the real part (χ′) observed in the data measured in the absence of magnetic-field are somewhat similar to those observed in the *dc* χ data. Additionally, the feature below 6 K shows a ν-dependence with noticeable imaginary component (χ″); it is to be noted that the peak in χ″ appears (near 6 K) only for the highest frequency employed (1.339 kHz), whereas at lower frequencies, just an upturn only could be observed, as though there is a peak below 2 K for such frequencies.

Viewing together all these data, it appears that the feature in the region 2 - 6 K is quite complex, attributable to spin-glass-like behavior (in the absence of an external magnetic field), particularly considering non-vanishing χ″ component [14].

The plot of *T*-dependence of dielectric permittivity (ε′) and loss factor (*tan δ*) in the absence of *dc* magnetic-field is shown in figure 3(a) at various frequencies. The results across $T_N$ were discussed in Ref. 5 and hence we restrict the present discussions to the data at further lower temperatures. For this reason, we have restricted the data in the mainframe to temperatures below 20 K, though we have shown the data in an extended *T*-range in the insets for the zero-field curves for one frequency. The values of *tanδ* are very low consistent with the insulating behavior of this compound, so that the observed features are not extrinsic. There is a change of curvature in the vicinity of 16-20 K, clearly visible for some frequencies (see insets), tracking magnetic anomalies in *C(T)* and χ(*T*), as though dielectric measurements are sensitive to crystal-field effects. A frequency dependent behavior could be seen in the data at low temperatures. A noteworthy feature is that there is a fall in ε′ below 8 K with a concomitant anomaly (a peak) in *tanδ*. It is worthwhile to note that the highest ν (1.339 kHz) employed in ac χ studies is close to the lowest ν (5 kHz) in dielectric studies and the temperature at which the anomaly is seen in both these data at these frequencies are somewhat close; as the ν is increased from 5 kHz to 100 kHz, the temperature at which the dielectric anomaly (say, a fall in ε′) occurs increases marginally to ~10 K. Considering these observations, one safely assume that the 10K-anomaly observed in this property is a consequence of the magnetic anomaly well below 6 K. The inset of figure 3(a) shows the plot of lnν versus inverse of *Tp*, fitted to Arrhenius law (ν= $\nu_0$ exp(-$E_a$/$k_B T_P$), $\nu_0$= pre-exponential factor, $E_a$= activation energy, $T_P$= peak temperature in χ″ ) and it is found that this plot is linear with an activation energy of 86 K. It was sufficiently stressed earlier [5] that dielectric anomalies track the features (<6K, ~16 K, ~ 32 K) in *C* and χ data, speaking in favor of ME coupling in this compound. Here we emphasize that the ν-dependence of the features in the same temperature range (below 10 K) both in ac χ and complex dielectric properties prompts us to infer that this compound exhibits 'reentrant multiglass-like' behavior, taking into account that the coupling occurring near $T_N$ is not of a glassy type, in analogy to the terminology applied for re-entrant spin-glasses. This observation is interesting, as this kind of 'reentrant mutiglass phenomenon' has been rarely reported for stoichiometric compounds, e.g., in $Dy_2BaNiO_5$ [3] and $Ca_3Co_2O_6$ [7]. As in the case of $Dy_2BaNiO_5$, an additional signature of spin-glass, viz., 'local dip' (or 'memory effects') in the *dc* χ experiments, performed with waiting and without waiting at a given temperature, could not be resolved down to 2 K and hence 'glassiness' is of a different origin, e.g., slow and complex dynamics arising from antiferromagnetically coupled Er and Ni magnetic moments.

**Influence of external dc magnetic fields on multiglass-like features:**

It is of interest to see how an application of external magnetic fields influences the magnetic and dielectric features. For this purpose, the data were collected while warming after cooling the sample in zero magnetic-field. It may be recalled [5] that, in the heat-capacity data taken in the presence of several magnetic fields, the temperature at which the peak occurs in *C/T* (around 16 K) gets lowered gradually with increasing magnetic fields. In the same manner, the corresponding broad peak (appearing around 16 K) in ac $\chi(T)$ data for *H*=0 (see figure 2) also shifts towards low temperatures with the application of magnetic field (for instance, with the fall in $\chi'$ occurring below around 5K for 30 kOe, see figure 2). As already remarked, these features are not strictly due to magnetic transitions, but due to an interplay between crystal-field effect and Zeeman Effect. The $\nu$-dependent component of $\chi''$ appearing below 6 K, which was proposed to arise from a magnetic anomaly at much lower temperatures also gets suppressed towards a lower temperature gradually (for instance, the peak in $\chi''$ for $\nu$= 1.339 kHz gets shifted from 6 K to 4 K as the *H* is increased to 30 kOe). Also, $\nu$-dependent behavior becomes less with the application of magnetic field (for instance, frequency dependence is ~ 1K for a variation from 1.3 Hz to 1.339 kHz for 30 kOe in $\chi''$). An application of magnetic field of 50 kOe further suppresses this low *T* feature, although the data is noisy for highest frequency.

Figure 3(b) and (c) show the $\nu$-dependent dielectric behavior under application of different magnetic fields. As discussed earlier [5] in favour of ME coupling, the peak temperature for a particular frequency shifts towards lower temperatures with increasing magnetic field, closely tracking the trend observed in $\chi''$ and *C(T)*. The $\nu$-dependence of the dielectric behavior gets gradually suppressed with increasing magnetic field, similar to that observed in ac $\chi$ (compare the plots in figures 2a and 2d with figures 3a and 3b). It is interesting to note that the $\nu$-dependent behavior of dielectric constant persists even in the presence of a field as high as 80 kOe, though the magnitude of the shift of characteristic temperature gets relatively reduced (not shown here). This $\nu$-dependent behavior almost vanishes by an application of a very high field of 140 kOe [Fig. 3c]. We could not measure ac $\chi$ above 50 kOe with our equipment for a comparative study at such high fields. Nevertheless, we can state that the tracking of the peak temperatures in ac $\chi$ and $\varepsilon'$ supports MDE coupling in this compound. It may be recalled that, for the compound $Dy_2BaNiO_5$, with increasing *H*, the low-temperature feature due to magnetism shifts towards higher *T*, whereas the one in dielectric shifts towards lower *T* [3]. This difference between Dy and Er compounds could be due to subtle differences in magnetic interactions and magnetic structure [9-11].

**A signature of magnetoelectric phase coexistence:**

Evidence for magnetodielectric coupling was obtained from the steps in the isothermal *M(H)* and $\varepsilon'(H)$ curves earlier [5]. The isothermal curves obtained by magnetic-field cycling reveals information about additional phenomenon, namely, phase-coexistence. The results of such measurements of $\varepsilon'$ as a function of *H* at different temperatures (with a rate of variation of 70 Oe/s) are presented in the form of magnetodielectric ($\Delta\varepsilon'= [(\varepsilon'_{H=0}- \varepsilon'_H)/\varepsilon'_{H=0}]$) effect in figure 4. In the virgin curve at 2K, there is a sudden upturn in the value around 20 kOe (and another weak one near 30 K), tracking the steps in *M(H)* (see also Ref. 5). It appears that these steps are due to disorder-broadened first-order transitions. As a consequence of this, the virgin curve should lie outside the envelope curve for a travel through a first-order transition for any phenomena [15], in the event that the high-field phase is retained at least partially on returning the magnetic field to zero, as observed experimentally in our case. It is thus interesting to note that $\varepsilon'$ in zero-field after the field-cycling attains a higher value compared to the virgin state zero-field value, as though there is an evidence for 'magnetoelectric phase co-existence' phenomenon. We could not resolve such a signature of 'magnetic phase-co-existence' in the isothermal magnetization data, although an extremely weak hysteresis (in the range 20 - 60 kOe) characteristic of first-order transition is observed around the metamagnetic transition fields (not shown here). Additional isothermal measurements were also performed in the following manner. The sample was magnetic-field-cooled from paramagnetic region (say, 200 K) to 2K with different magnetic fields and then the field was removed; the isothermal field-dependent dielectric measurements were then performed with a sweep rate of 70 Oe/min (Fig. 4(b) and 4(c)). It is found that an application of magnetic field of 50 kOe decreases the degree of partial arrest and this phase-coexistence phenomenon vanishes by an application of very high fields (say, 140 kOe). These peculiar features in dielectric (without similar anomalies in isothermal magnetization) has been observed by us in another compound $Ca_3Co_2O_6$ [7], thereby suggesting that this discrepancy should be more common, thereby tempting the community to search for similar materials to understand the dynamics associated with dielectric phase coexistence phenomenon. The feature attributable to phase-coexistence phenomenon is not observed at higher temperatures (above 5 K) [see Fig. 4d for T= 8K], clearly indicating a change in the magnetic state across these temperatures.

**CONCLUSIONS**

We have investigated the Haldane spin-chain compound, $Er_2BaNiO_5$, at temperatures below its long-range antiferromagnetic ordering temperature of 32 K by *ac* and *dc* magnetization as well as dielectric studies. Apart from the fact that the dielectric property captures the features due to the crystal-field effects proposed in the literature, the results provide evidence for another magnetic anomaly below 6 K. It turns out that the 6K-anomaly characterizes this compound 'reentrant multi-glass-like', based on frequency dependence of ac susceptibility and dielectric properties. As a manifestation of metamagnetic transition, 'magnetoelectric phase coexistence' is observed at very low temperatures.

**Acknowledgments**

The authors thank Kartik K Iyer for his valuable help while carrying out experiments.

Figure 1:
*Dc* magnetization divided by magnetic-field as a function of temperature (1.8 - 40K) obtained in 100 Oe for zero-field-cooled and field-cooled conditions of the specimen. Inset (a) shows the derivative curve to highlight the feature at $T_N$. Inset (b) shows isothermal remnant magnetization as a function of time (*t*) at 1.8K, obtained as described in the text.

Figure 2:
Real ($\chi'$) and imaginary ($\chi''$) parts of *ac* susceptibility as a function of temperature below 20 K in 0, 10, 30, and 50 kOe, obtained with various frequencies. In the inset, the zero-field data over an extended *T*-range is shown to highlight the broad peak around 16 K.

Figure 3:
 (a) Dielectric constant ($\varepsilon'$) and the loss factor (tan$\delta$) as a function of temperature below 20 K (in the absence of an external magnetic field), obtained with various frequencies. The curves obtained at higher fields are shown in (b) and (c) for 50 and 140 kOe respectively. The arrows are shown to identify the curves with increasing frequency. In the insets of (a) and (b), the curves for 5 kHz are shown over an extended *T*-range. In another inset of (a), Arrhenius plot is shown.

Figure 4:
The fractional change in dielectric constant as a function of magnetic field (a) at 2 K for zero-field-cooled (ZFC) condition, (b) at 2 K for field-cooled (FC) from 200K with *H*= 50kOe, (c) at 2 K for field-cooled from 200K with H=140 kOe, and (d) at 8 K for ZFC condition, obtained with a frequency of 50 kHz. Arrows and numericals are placed to show the direction of magnetic-field variation from the starting point. The plots around zero-field region are shown in an expanded form separately in the insets in (a) and (d).

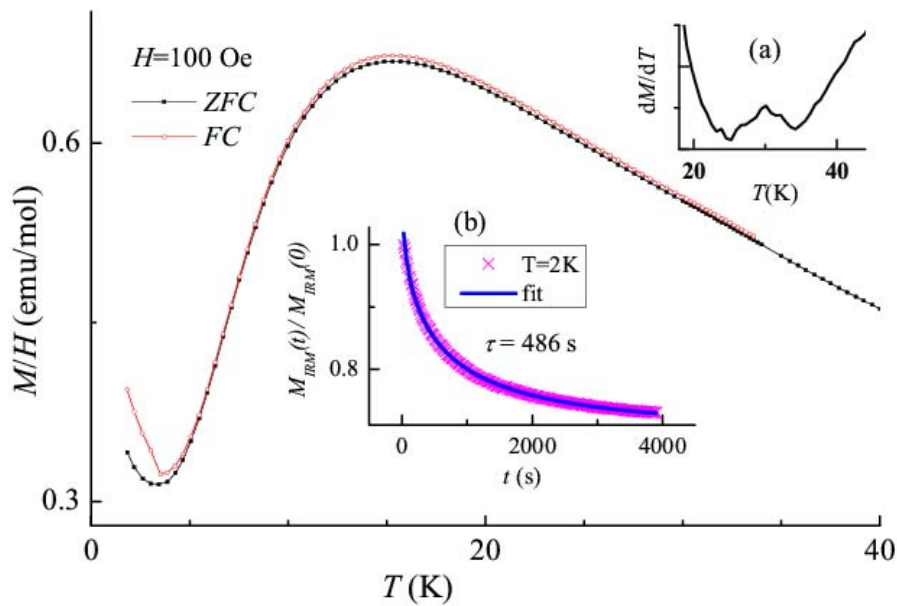

Figure 1:

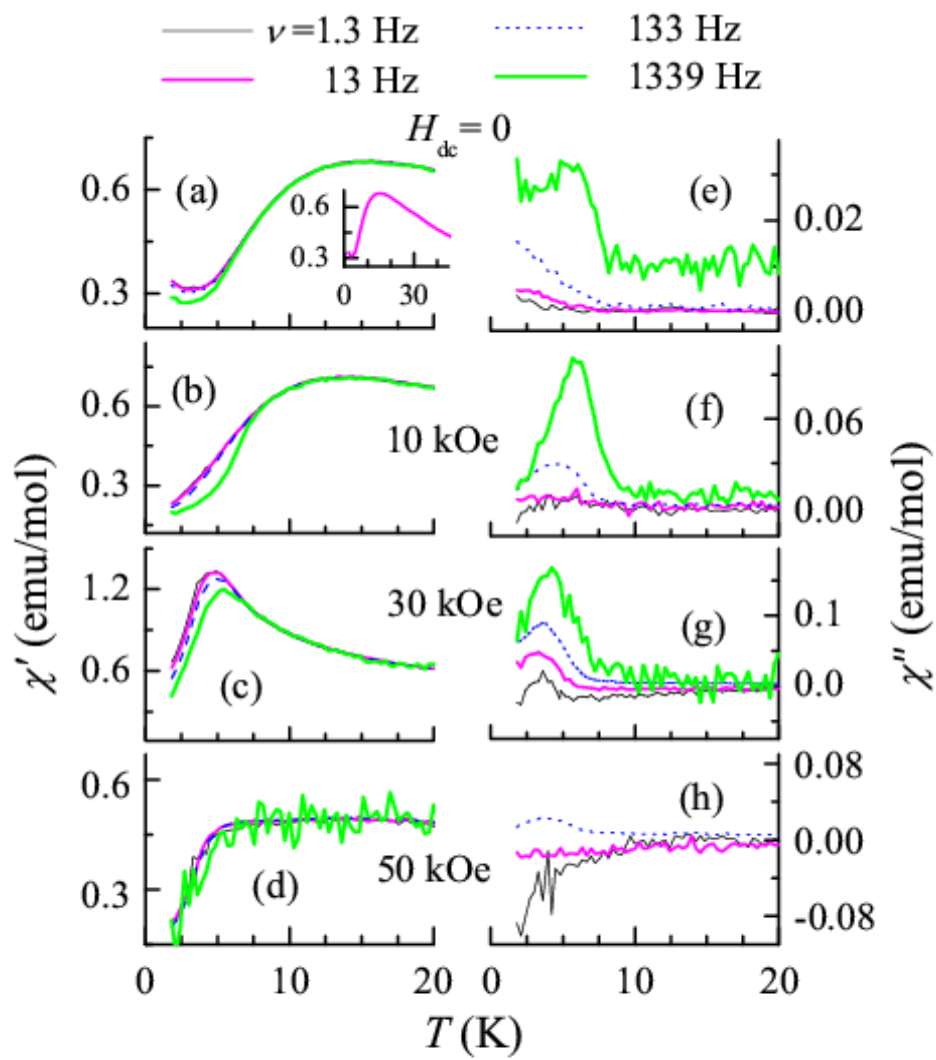

Figure 2:

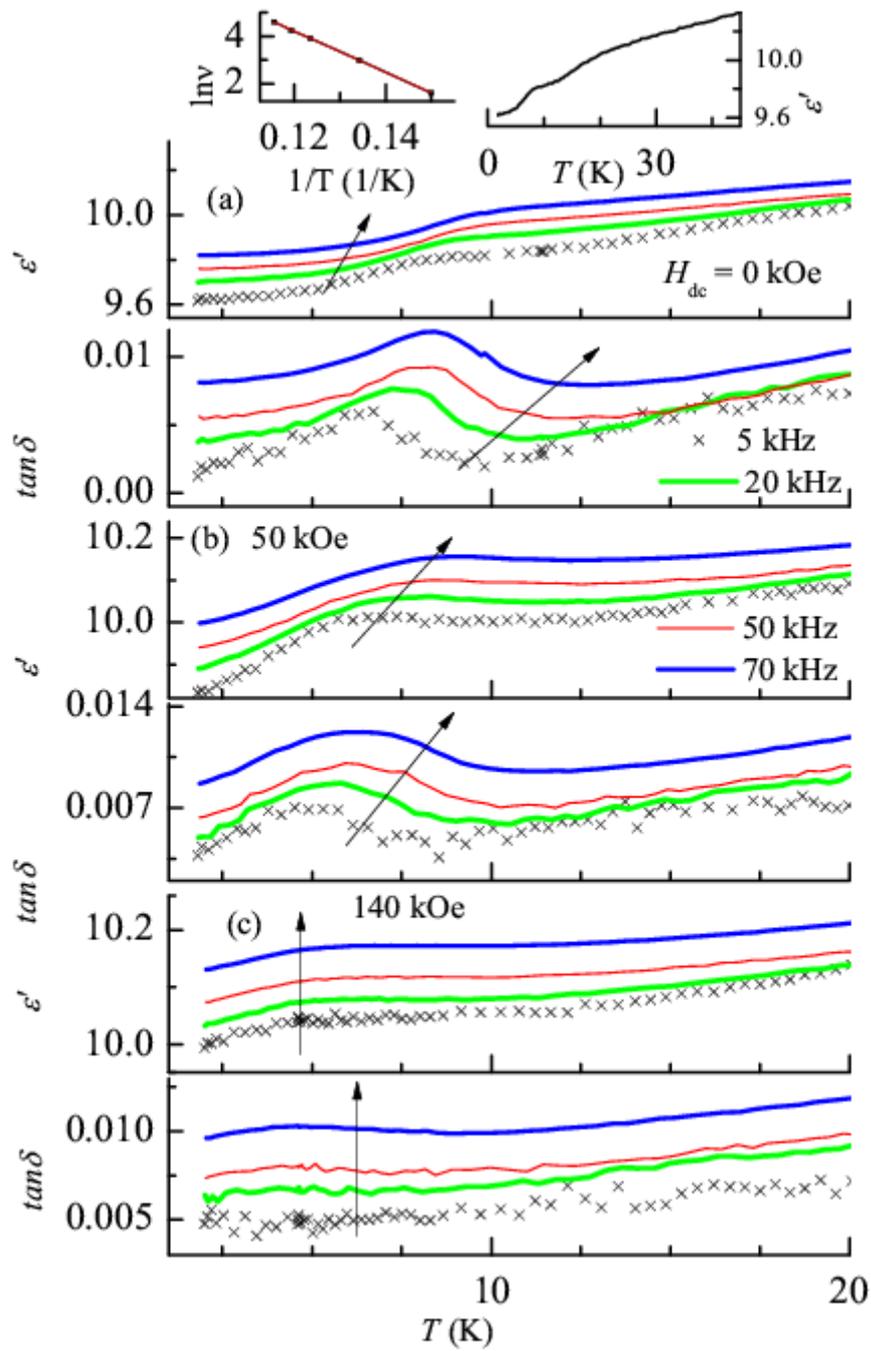

Figure 3

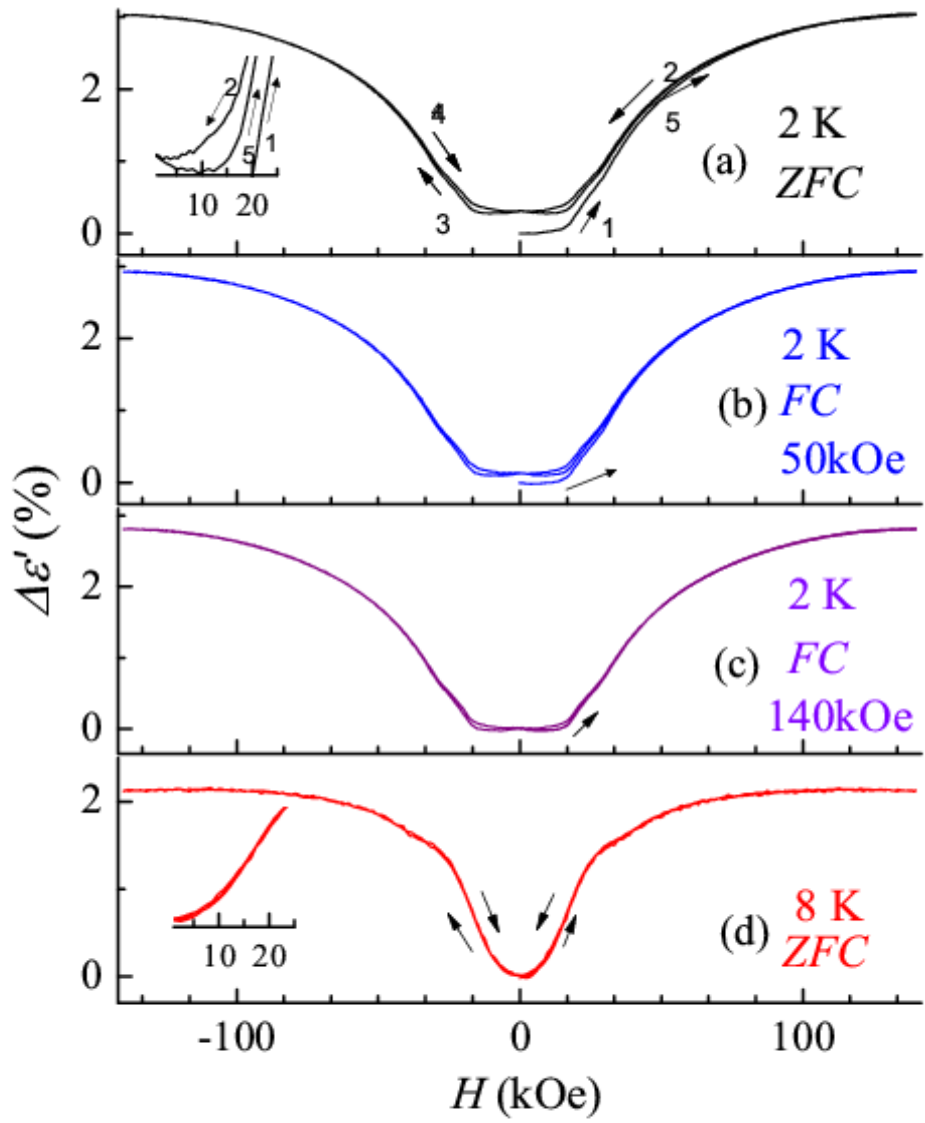

Figure 4